# A High Performance Disturbance Observer: A Unified Synthesis Approach

Emre Sariyildiz
School of Mechanical, Materials, Mechatronic, and Biomedical Engineering,
Faculty of Engineering and Information Sciences
University of Wollongong
Wollongong, NSW, 2522, Australia
emre@uow.edu.au

*Abstract*— This paper presents a new High-Performance Disturbance Observer (HPDOb) that significantly improves disturbance estimation accuracy and robustness in motion control systems, surpassing the capabilities of conventional disturbance observers (DObs). Both conventional and proposed observers are analysed and synthesised in the discrete-time domain, providing a realistic representation of their dynamic behaviour and enabling enhanced controller design for practical applications. The core contribution of the HPDOb is a novel synthesis method that incorporates higher-order truncation error dynamics into disturbance estimation. Unlike conventional DObs, which are limited to zero-order truncation error, the HPDOb achieves first-order truncation error, yielding markedly improved estimation accuracy and robustness against disturbances in motion control systems. Furthermore, the framework supports flexible error-dynamics design, including higher-order truncation models, thereby broadening its applicability to a wide range of robust motion control problems. Simulation and experimental studies validate that the HPDOb substantially enhances disturbance estimation and control performance in both regulation and trajectory tracking tasks under internal and external disturbances of servo systems.

## I. INTRODUCTION

Over the past three decades, Disturbance Observers (DObs) have become a cornerstone of motion control research, particularly in the development of robust control strategies [1–6]. DOb-based controllers have been applied across diverse engineering domains, ranging from high-precision position control of servo mechanisms to force estimation in robotic systems and compensation for communication delays [6–13]. In these frameworks, robustness is achieved by embedding disturbance estimates within an inner control loop, thereby mitigating the impact of uncertainties and external disturbances [13 –15]. To enable effective implementation of this strategy, a wide variety of DOb synthesis techniques have been proposed and extensively studied [15–20].

The synthesis of a DOb requires defining a dynamic model of the disturbances acting on the servo system [21]. Conventional DObs are typically designed in the continuous-time domain under the simplifying assumption that disturbances remain constant [1, 13]. However, this traditional approach entails two major limitations. First, since robust motion control systems are almost always implemented on digital platforms such as microcontrollers and computers, continuous-time models fail to fully capture their discrete-time behaviour, which can lead to unforeseen stability and performance issues [5, 23, 24]. Second, the assumption of constant disturbances is rarely valid in practice, as servo systems are frequently subject to time-varying disturbances [26, 27]. This oversimplification restricts estimation accuracy, thereby constraining robustness and ultimately degrading control performance in real-world scenarios [28].

In the quest for improved robust stability and performance, numerous approaches for analysing and synthesising digital DOb-based robust motion controllers have been proposed [22–33]. For example, robust stability and performance analyses of digital DObs were presented in [30, 31], while optimal tuning methods for controller parameters were introduced in [32, 33]. Nevertheless, the accuracy of disturbance estimation remains fundamentally constrained by the conventional assumption of constant disturbances in DOb synthesis. To overcome this limitation, researchers have explored alternative disturbance models. Periodic DObs, for instance, can effectively address periodic disturbances [34], while generalised DObs employing higher-order disturbance estimation offer further improvements in accuracy [35, 36]. However, higher-order DObs tend to be more sensitive to measurement noise [35–37], leading to a challenging trade-off between estimation accuracy and noise robustness, which is a well-known trade-off that becomes particularly critical when such observers are integrated into robust motion controllers [15, 19, 35]. More recently, a high-performance DOb based on a predictor–observer synthesis method was introduced [38], enhancing disturbance estimation through a more realistic disturbance model. Yet, this method departs from the conventional DOb synthesis framework, limiting its general applicability [39]. These developments highlight the need for digital DObs that achieve both enhanced robustness and improved estimation accuracy, while maintaining compatibility with established synthesis principles.

This paper introduces a novel DOb synthesis approach in the discrete-time domain that achieves superior disturbance estimation accuracy and enhanced robustness for motion control applications. Unlike conventional DObs, the proposed HPDOb incorporates higher-order Taylor series approximations of disturbances into the synthesis process, thereby significantly improving estimation accuracy [40, 41]. This advancement translates into notable benefits for motion control tasks, including increased robustness in both position regulation and trajectory tracking. The effectiveness of the proposed HPDOb is verified through comprehensive simulations and experimental studies, which demonstrate that the improved disturbance estimation directly enables enhanced control performance under internal and external disturbances in servo systems.

The rest of the paper is organised as follows. Section II reviews the conventional disturbance observer synthesis within the discrete-time domain, presenting a straightforward yet effective design approach for robust motion control applications. Section III introduces the proposed High-Performance Disturbance Observer (HPDOb), which offers improved disturbance estimation accuracy compared to the conventional method. Section IV validates the effectiveness of the HPDOb through simulations and experiments. Finally, Section V concludes the paper.

## II. DOb Synthesis in Motion Control

The continuous-time state-space dynamics of a servo-based motion control system can be formulated as:

$$\dot{\mathbf{x}}(t) = \mathbf{A}_\mathbf{C}\mathbf{x}(t) + \mathbf{B}_\mathbf{C}u(t) - \mathbf{D}_\mathbf{C}\tau_d(t) \tag{1}$$

Here, $\mathbf{x}(t)$ denotes the state vector, with position and velocity appearing in the first and second entries, respectively. The system dynamics are captured by the state matrix $\mathbf{A}_\mathbf{C}$, whose first row is [0 1], and whose second row contains [0 $-b_m/J_m$], reflecting the ratio of the viscous friction coefficient $b_m$ to the motor inertia $J_m$. The exogenous control and disturbance input signals are $u(t)$ and $\tau_d(t)$, and they integrate into the dynamic model through control input and disturbance vectors $\mathbf{B}_\mathbf{C}$ and $\mathbf{D}_\mathbf{C}$ whose first and second entries are 0 and $1/J_m$.

A corresponding nominal dynamic model can be defined using Eq. (2) as follows:

$$\dot{\mathbf{x}}(t) = \mathbf{A}_\mathbf{Cn}\mathbf{x}(t) + \mathbf{B}_\mathbf{Cn}u(t) - \mathbf{D}_\mathbf{Cn}\tau_{dn}(t) \tag{2}$$

where $\mathbf{A}_\mathbf{Cn}$, $\mathbf{B}_\mathbf{Cn}$ and $\mathbf{D}_\mathbf{Cn}$ are the nominal state and input matrices, determined from the nominal inertia $J_{mn}$ and viscous friction coefficient $b_{mn}$. Using Eqs. (1) and (2), the nominal disturbance variable can be expressed as

$$\tau_{dn}(t) = \frac{\mathbf{D}_\mathbf{Cn}^T}{\mathbf{D}_\mathbf{Cn}^T\mathbf{D}_\mathbf{Cn}}\left(\Delta\mathbf{A}_\mathbf{C}\mathbf{x}(t) + \Delta\mathbf{B}_\mathbf{C}u(t) + \mathbf{D}_\mathbf{C}\tau_d(t)\right) \tag{3}$$

where $\Delta\mathbf{A}_\mathbf{C} = \mathbf{A}_\mathbf{Cn} - \mathbf{A}_\mathbf{C}$ and $\Delta\mathbf{B}_\mathbf{C} = \mathbf{B}_\mathbf{Cn} - \mathbf{B}_\mathbf{C}$ represent the parametric uncertainties in the dynamic model.

Equations (1–3) provide a compact and simple yet effective description of the servo dynamics used in robust motion control analysis. For further modelling details in DOb-based robust motion control, the reader is referred to [2, 27].

The state-space dynamics of the servo system can also be expressed in the discrete-time domain using Eq. (4).

$$\mathbf{x}_{k+1} = \mathbf{A}_\mathbf{D}\mathbf{x}_k + \mathbf{B}_\mathbf{D}u_k - \mathbf{\Pi}_{\mathbf{D}_k} \tag{4}$$

where $\mathbf{A}_\mathbf{D} = e^{\mathbf{A}_\mathbf{C}T_s}$ is the discrete-time state matrix; $\mathbf{B}_\mathbf{D} = \int_0^{T_s} e^{\mathbf{A}_\mathbf{C}\tau}\mathbf{B}_\mathbf{C}d\tau$ and $\mathbf{\Pi}_{\mathbf{D}_k} = \int_0^{T_s} e^{\mathbf{A}_\mathbf{C}\tau}\mathbf{D}_\mathbf{C}\tau_d\left((k+1)T_s - \tau\right)d\tau$ are the discrete-time control and disturbance input vectors, respectively; $\mathbf{x}_k$ is the state vector denoting the servo system's position and velocity at the sampling instant $kT_s$; and $u_k$ is the control input at the same instant [38].

Similarly, the nominal state-space model of the servo system in the discrete-time domain is as follows:

$$\mathbf{x}_{k+1} = \mathbf{A}_\mathbf{Dn}\mathbf{x}_k + \mathbf{B}_\mathbf{Dn}u_k - \mathbf{\Pi}_{\mathbf{Dn}_k} \tag{5}$$

where $\mathbf{A}_\mathbf{Dn} = e^{\mathbf{A}_\mathbf{Cn}T_s}$ is the nominal discrete-time state matrix, and $\mathbf{\Pi}_{\mathbf{Dn}_k} = \int_0^{T_s} e^{\mathbf{A}_\mathbf{Cn}\tau}\mathbf{D}_\mathbf{Cn}\tau_{dn}\left((k+1)T_s - \tau\right)d\tau$ and $\mathbf{B}_\mathbf{Dn} = \int_0^{T_s} e^{\mathbf{A}_\mathbf{Dn}\tau}\mathbf{B}_\mathbf{Dn}d\tau$ are the nominal discrete-time disturbance and control input vectors, respectively [24].

To synthesise a DOb, we need to introduce a nominal model for the disturbance variable. Let us derive the dynamic model of the disturbance vector using Eq. (6).

$$\mathbf{\Pi}_{\mathbf{Dn}_k} = \int_0^{T_s} e^{\mathbf{A}_\mathbf{Cn}\lambda}\mathbf{D}_\mathbf{Cn}\left(\sum_{i=0}^{m}\left(\frac{1}{i!}\tau_{dn_k}^{(i)}(T_s - \lambda)^i\right) + R_{m_k}\right)d\lambda \tag{6}$$

where $R_{m_k} = O(m+1)$ represents the $(m+1)^{th}$ order truncation error, and $\tau_{dn}^{(i)}$ represents the $i^{th}$ order derivative of $\tau_{dn}$.

To derive the conventional DOb, the truncation error is assumed to be zero, i.e., $R_{0_k} = O(1) = \dot{\tau}_{dn_{\xi_k}}(T_s - \lambda) = 0$ where $\xi_k$ is an unknown time within the sampling period. Under this assumption, the nominal disturbance is treated as constant over each sampling interval, so we can use the following nominal disturbance model in the DOb synthesis.

$$\tau_{dn_{\xi_k}} = \tau_{dn}(kT_s) \quad \text{when} \quad kT_s \leq \xi_k < (k+1)T_s \tag{7}$$

Substituting (7) into the nominal discrete-time model (5) yields a simplified dynamic representation of the servo system given by

$$\mathbf{x}_{k+1} = \mathbf{A}_\mathbf{Dn}\mathbf{x}_k + \mathbf{B}_\mathbf{Dn}u_k - \mathbf{D}_\mathbf{Dn}\tau_{dn_k} \tag{8}$$

where $\mathbf{D}_\mathbf{Dn} = \int_0^{T_s} e^{\mathbf{A}_\mathbf{Cn}\tau}\mathbf{D}_\mathbf{Cn}d\tau$ and $\tau_{dn_k}$ is the pseudo-disturbance acting on the nominal system at the sampling instant $kT_s$. This formulation provides the basis for conventional DOb synthesis in robust motion control.

The conventional DOb in the discrete-time domain can be synthesised using the auxiliary variable defined as

$$z_k = \tau_{dn_k} + \mathbf{L}^T\mathbf{x}_k \tag{9}$$

where $z_k \in R$ is the auxiliary variable, which is constructed by summing the nominal disturbance variable and the state vector multiplied by the observer gain vector $\mathbf{L}$ at the sampling instant $kT_s$.

Since the nominal disturbance variable is unknown at $kT_s$ seconds, we can't directly measure the auxiliary variable. An observer that can estimate the auxiliary variable can be synthesised by deriving its discrete-time dynamic model as follows:

$$z_{k+1} = \tau_{dn_{k+1}} + \mathbf{L}^T \mathbf{x}_{k+1} = \tau_{dn_k} + \mathbf{L}^T \mathbf{x}_{k+1} + \tau_{dn_{k+1}} - \tau_{dn_k}$$
$$= (1 - \mathbf{L}^T \mathbf{D_{Dn}}) z_k + \mathbf{L}^T (\mathbf{A_{Dn}} + \mathbf{D_{Dn}} \mathbf{L}^T - \mathbf{I_2}) \mathbf{x}_k + \mathbf{L}^T \mathbf{B_{Dn}} u_k + \tau_{dn_{k+1}} - \tau_{dn_k} \quad (10)$$

where $\mathbf{I_2}$ is an identity matrix.

Using Eq. (10), an observer for the auxiliary variable is synthesised as follows:

$$\hat{z}_{k+1} = (1 - \mathbf{L}^T \mathbf{D_{Dn}}) \hat{z}_k + \mathbf{L}^T (\mathbf{A_{Dn}} + \mathbf{D_{Dn}} \mathbf{L}^T - \mathbf{I_2}) \mathbf{x}_k + \mathbf{L}^T \mathbf{B_{Dn}} u_k \quad (11)$$

where $\hat{z}_k$ is the estimated $z_k$ at $kT_s$ seconds.

Subtracting Eq. (11) from Eq. (10) yields

$$z_{k+1} - \hat{z}_{k+1} = (1 - \mathbf{L}^T \mathbf{D_{Dn}})(z_k - \hat{z}_k) + \tau_{dn_{k+1}} - \tau_{dn_k} \quad (12)$$

Equation (12) shows that the estimation error of the auxiliary variable is bounded by the variation of the disturbance within the sampling period, provided that the observer gain is properly tuned $|1 - \mathbf{L}^T \mathbf{D_{Dn}}| < 1$. In other words, the conventional disturbance observer is uniformly ultimately bounded (UUB), indicating an upper bound on the estimation error and the stability of the observer. When the disturbance variable is constant within the sampling period, the estimation error converges to zero, so asymptotic stability can be achieved. The nominal disturbance variable can be estimated using Eqs. (9) and (11) as follows:

$$\hat{\tau}_{dn_k} = \hat{z}_k - \mathbf{L}^T \mathbf{x}_k \quad (13)$$

where $\hat{\tau}_{dn_k}$ is the estimated nominal disturbance variable at the sampling instant $kT_s$.

### III. A UNIFIED HP-DOB SYNTHESIS IN MOTION CONTROL

As shown in Eqs. (6), (7) and (12), the performance of the conventional DOb is limited by the truncation error of order 0. Let us improve the performance of disturbance estimation by obtaining a higher-order truncation error in the synthesis of DOb.

To synthesise a high-performance DOb, let us consider the following auxiliary variables.

$$z_{1_k} = \tau_{dn_{k-1}} + \mathbf{L_1}^T \mathbf{x}_k$$
$$z_{2_k} = \tau_{dn_k} + \mathbf{L_2}^T \mathbf{x}_k \quad (14)$$

Let us first consider the auxiliary variable $z_{1_k}$. Its dynamic equations can be obtained as follows:

$$z_{1_{k+1}} = \tau_{dn_k} + \mathbf{L_1}^T \mathbf{x}_{k+1} = \tau_{dn_k} + \mathbf{L_1}^T (\mathbf{A_{Dn}} \mathbf{x}_k + \mathbf{B_{Dn}} u_k - \mathbf{D_{Dn}} \tau_{dn_k}) \quad (15)$$

Substituting $\tau_{dn_k} = z_{2_k} - \mathbf{L_2}^T \mathbf{x}_k$, which is derived from Eq. (14), into Eq. (15) yields

$$z_{1_{k+1}} = z_{2_k} - \mathbf{L_2}^T \mathbf{x}_k + \mathbf{L_1}^T (\mathbf{A_{Dn}} \mathbf{x}_k + \mathbf{B_{Dn}} u_k - \mathbf{D_{Dn}} (z_{2_k} - \mathbf{L_2}^T \mathbf{x}_k))$$
$$= (1 - \mathbf{L_1}^T \mathbf{D_{Dn}}) z_{2_k} + \mathbf{L_1}^T \mathbf{B_{Dn}} u_k + (\mathbf{L_1}^T \mathbf{A_{Dn}} + \mathbf{L_1}^T \mathbf{D_{Dn}} \mathbf{L_2}^T - \mathbf{L_2}^T) \mathbf{x}_k \quad (16)$$

Let us now consider the second auxiliary variable $z_{2_k}$. Its dynamic equations can be similarly obtained as follows:

$$z_{2_{k+1}} = \tau_{dn_{k+1}} + \mathbf{L_2}^T \mathbf{x}_{k+1} = 2\tau_{dn_k} + \mathbf{L_2}^T \mathbf{x}_{k+1} + \tau_{dn_{k+1}} - 2\tau_{dn_k}$$
$$= 2\tau_{dn_k} + \mathbf{L_2}^T (\mathbf{A_{Dn}} \mathbf{x}_k + \mathbf{B_{Dn}} u_k - \mathbf{D_{Dn}} \tau_{dn_k}) + \tau_{dn_{k+1}} - 2\tau_{dn_k} \quad (17)$$

Substituting $\tau_{dn_k} = z_{2_k} - \mathbf{L_2}^T \mathbf{x}_k$, which is derived from Eq. (14), into Eq. (17) yields

$$z_{2_{k+1}} = (2 - \mathbf{L_2}^T \mathbf{D_{Dn}}) z_{2_k} + \mathbf{L_2}^T \mathbf{B_{Dn}} u_k + \mathbf{L_2}^T (\mathbf{A_{Dn}} + \mathbf{D_{Dn}} \mathbf{L_2}^T - 2\mathbf{L_2}^T) \mathbf{x}_k + \tau_{dn_{k+1}} - 2\tau_{dn_k} \quad (18)$$

Using $\tau_{dn_{k-1}} = z_{1_k} - \mathbf{L_1}^T \mathbf{x}_k$ from Eq. (14), Eq. (18) can also be written as follows:

$$z_{2_{k+1}} = -\tau_{dn_{k-1}} + (2 - \mathbf{L_2}^T \mathbf{D_{Dn}}) z_{2_k} + \mathbf{L_2}^T \mathbf{B_{Dn}} u_k +$$
$$\mathbf{L_2}^T (\mathbf{A_{Dn}} + \mathbf{D_{Dn}} \mathbf{L_2}^T - 2\mathbf{L_2}^T) \mathbf{x}_k + \tau_{dn_{k+1}} - 2\tau_{dn_k} + \tau_{dn_{k-1}}$$
$$= -z_{1_k} + (2 - \mathbf{L_2}^T \mathbf{D_{Dn}}) z_{2_k} + \mathbf{L_2}^T \mathbf{B_{Dn}} u_k +$$
$$(\mathbf{L_1}^T + \mathbf{L_2}^T \mathbf{A_{Dn}} + \mathbf{L_2}^T \mathbf{D_{Dn}} \mathbf{L_2}^T - 2\mathbf{L_2}^T) \mathbf{x}_k + \tau_{dn_{k+1}} - 2\tau_{dn_k} + \tau_{dn_{k-1}} \quad (19)$$

Using Eqs. (16) and (19), the proposed HP-DOb can be synthesised as follows:

$$\hat{z}_{1_{k+1}} = (1 - \mathbf{L_1}^T \mathbf{D_{Dn}}) \hat{z}_{2_k} + \mathbf{L_1}^T \mathbf{B_{Dn}} u_k + (\mathbf{L_1}^T \mathbf{A_{Dn}} + \mathbf{L_1}^T \mathbf{D_{Dn}} \mathbf{L_2}^T - \mathbf{L_2}^T) \mathbf{x}_k \quad (20)$$

$$\hat{z}_{2_{k+1}} = -\hat{z}_{1_k} + (2 - \mathbf{L_2}^T \mathbf{D_{Dn}}) \hat{z}_{2_k} + \mathbf{L_2}^T \mathbf{B_{Dn}} u_k + (\mathbf{L_1}^T + \mathbf{L_2}^T \mathbf{A_{Dn}} + \mathbf{L_2}^T \mathbf{D_{Dn}} \mathbf{L_2}^T - 2\mathbf{L_2}^T) \mathbf{x}_k \quad (21)$$

Subtracting Eqs. (20) and (21) from Eqs. (16) and (19) yields the estimation error dynamics as

$$\begin{bmatrix} z_{1_{k+1}} - \hat{z}_{1_{k+1}} \\ z_{2_{k+1}} - \hat{z}_{2_{k+1}} \end{bmatrix} = \begin{bmatrix} 0 & 1 - \mathbf{L_1}^T \mathbf{D_{Dn}} \\ -1 & 2 - \mathbf{L_2}^T \mathbf{D_{Dn}} \end{bmatrix} \begin{bmatrix} z_{1_k} - \hat{z}_{1_k} \\ z_{2_k} - \hat{z}_{2_k} \end{bmatrix} + \begin{bmatrix} 0 \\ \tau_{dn_{k+1}} - 2\tau_{dn_k} + \tau_{dn_{k-1}} \end{bmatrix} \quad (22)$$

Similarly, Eq. (22) shows that the estimation errors of the auxiliary variables are bounded by the modified disturbance variation variable $(\tau_{dn_{k+1}} - 2\tau_{dn_k} + \tau_{dn_{k-1}})$ within the sampling period, provided that the observer gains are properly tuned so that the eigenvalues of the discrete-time estimation error matrix $\left( \begin{bmatrix} 0 & 1 - \mathbf{L_1}^T \mathbf{D_{Dn}} \\ -1 & 2 - \mathbf{L_2}^T \mathbf{D_{Dn}} \end{bmatrix} \right)$ are smaller than 1. Since the eigenvalues of the error matrix can be independently assigned, we can obtain uniformly ultimately bounded estimation error when the nominal disturbance variable is bounded at sampling instances. When the modified disturbance estimation variable is zero, the estimation error goes to zero and asymptotic stability is similarly achieved for the proposed HP-DOb.

Using Eqs. (14) and (22), the estimated disturbance is obtained as follows:

$$\hat{\tau}_{dn_k} = \hat{z}_{2_k} - \mathbf{L}_2^T \mathbf{x}_k \quad (23)$$

where $\hat{\tau}_{dn_k}$ is similarly the estimated nominal disturbance variable at the sampling instant $kT_s$, however it is derived using the second auxiliary variable.

The dynamics of the disturbance estimation error can be adjusted by tuning the eigenvalues of the discrete-time estimation error state matrix through the observer gains $\mathbf{L}_1$ and $\mathbf{L}_2$. For example, the observer gains of the HP-DOb can be tuned using Eq. (24).

$$\begin{aligned}\mathbf{L}_1^T \mathbf{D}_{\mathbf{Dn}} &= 1 - \lambda_1^{des} \lambda_2^{des} \\ \mathbf{L}_2^T \mathbf{D}_{\mathbf{Dn}} &= 2 - \lambda_1^{des} - \lambda_2^{des}\end{aligned} \quad (24)$$

where $\lambda_1^{des}$ and $\lambda_2^{des}$ are the desired eigenvalues of the discrete-time estimation error state matrix.

Compared with the error dynamics of the conventional DOb in Eq. (12), the error dynamics of the proposed HP-DOb incorporate a first-order truncation error term. As a result, the HP-DOb achieves higher disturbance estimation accuracy than the conventional design.

## IV. SIMULATIONS

This section presents simulation and experimental studies conducted to evaluate the stability and performance of a robust position control system employing both the conventional and the proposed high-performance DObs. The DObs are implemented in the inner loop to suppress disturbances, while a Proportional–Derivative (PD) controller is applied in the outer loop to regulate motion performance. For consistency between simulations and experiments, the outer-loop controller parameters were fixed, with the proportional and derivative gains set to Kp = 50 and Kd = 5, respectively. The experimental setup is shown in Fig. 1, which we used Macon EC90: 607931 flat motor for position control and applying external disturbances. They were controlled using Maxon's Escon motor driver and a Matlab/Simulink real-time controller with the sampling period 1ms.

The observer gains of the conventional and proposed high-performance DObs were tuned using Eq. (25).

$$\mathbf{L}_1 = \frac{k_1}{\left|\mathbf{D}_{\mathbf{Dn}}\right|_1}\begin{bmatrix}1 & 1\end{bmatrix}^T \text{ and } \mathbf{L}_2 = \frac{k_2}{\left|\mathbf{D}_{\mathbf{Dn}}\right|_1}\begin{bmatrix}1 & 1\end{bmatrix}^T \quad (25)$$

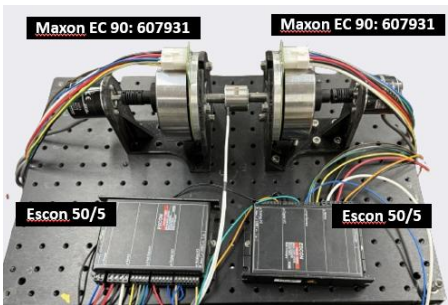

Fig. 1: Experimental setup.

where $k_1$ and $k_2$ are free control parameters tuned by using Eqs. (12) and (24) in the conventional and high-performance DOb synthesis, respectively; and $\left|\mathbf{D}_{\mathbf{Dn}}\right|_1$ is the L1-norm of $\mathbf{D}_{\mathbf{Dn}}$.

Let us begin with position control simulations. Figure 2a shows the position response of the PD-controlled servo system under the external disturbances depicted in Fig. 2b. As seen in Fig. 2, the presence of disturbances leads to a significant degradation in position control performance when only the PD controller is employed. Therefore, robust motion control techniques are essential to achieve high-performance when servo systems are affected by disturbances.

To enhance robustness against disturbances, the conventional DOb is implemented in the inner loop, while the same PD controller is retained in the outer loop. The resulting position control performance is presented in Fig. 3a. In comparison with Fig. 2, the conventional DOb significantly improves the system's disturbance rejection capability in both regulation and trajectory tracking control. Although the disturbance profile remains similar, as shown in Fig. 3b, the estimation accuracy improves with an increase in the DOb bandwidth (Fig. 3c). However, the bandwidth cannot be increased arbitrarily, as stability and performance limitations impose constraints on this design parameter. For a detailed discussion of the stability–performance trade-off in DOb-based motion control systems, the reader is referred to [15].

Figure 4 compares the position control performance of the servo system when the conventional DOb and the proposed HPDOb are employed in the inner loop and the same performance controller is employed in the outer-loop. Both observers can estimate disturbances accurately and enhance

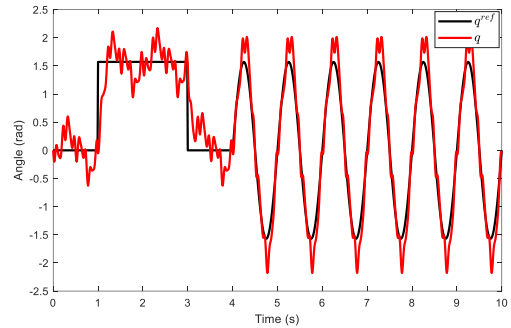

a)   Position control of the servo system.

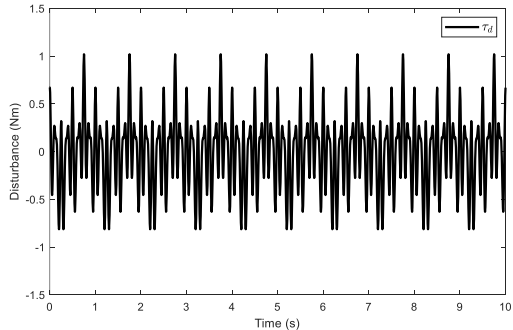

b)   Disturbances acting on the servo system.

Fig. 2: Position control using PD controller.

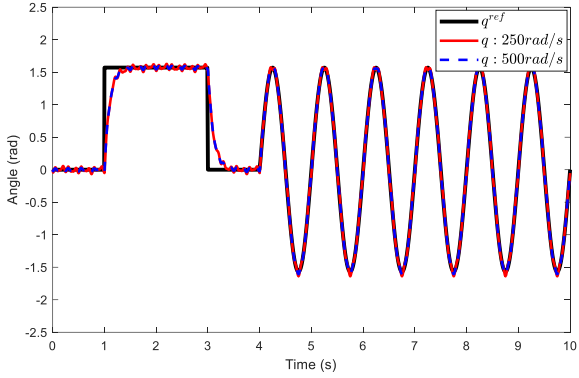

a) Position control of the servo system.

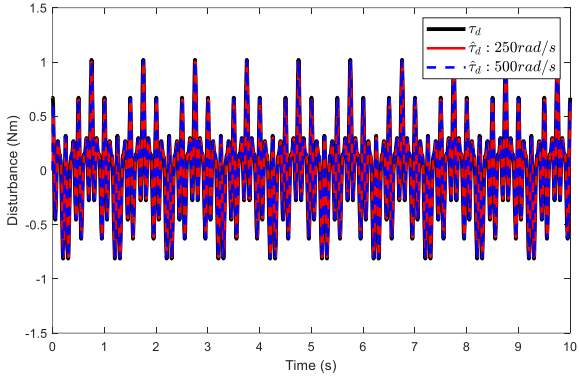

b) Disturbances acting on the servo system and their estimates.

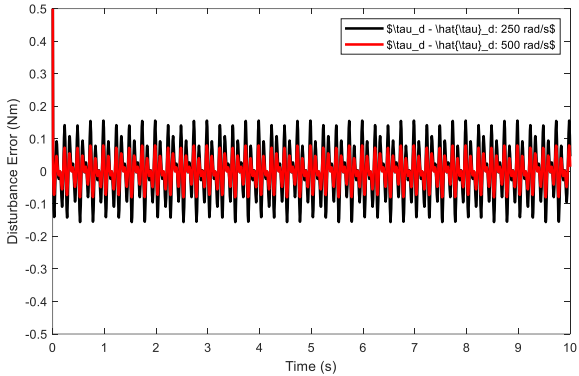

c) Disturbance estimation error.

Fig. 3: Position control using PD controller and conventional DObs.

robustness against disturbances and enable accurate regulation and trajectory tracking as shown in Figs. 4a. However, the HPDOb-based controller achieves superior performance compared with the conventional design, owing to its more accurate disturbance estimation, as shown in Fig. 4b.

Figure 5 illustrates the position control experiment of the HPDOb-based robust motion controller. As shown in this figure, the proposed DOb can accurately estimate external disturbances, thereby enabling high-performance motion control applications.

## V. CONCLUSION

This paper has presented a novel disturbance observer (DOb). Unlike conventional DObs, which rely on a nominal disturbance model with zero-order truncation error, the proposed HPDOb incorporates higher-order truncation error dynamics into the observer synthesis. This approach significantly improves the accuracy of disturbance estimation, thereby enhancing overall system performance. Simulation and experimental results confirm that the HPDOb enables precise tracking of both set-point and trajectory references by improving robustness against disturbances through more accurate estimation. Beyond disturbance rejection, the framework provides a flexible foundation for designing observers with tailored error dynamics, opening new possibilities for advanced robust control architectures. Future studies should further investigate the robust stability and performance of HPDOb-based controllers and explore their integration into force and impedance control applications, as well as broader domains such as robotics and precision mechatronics.

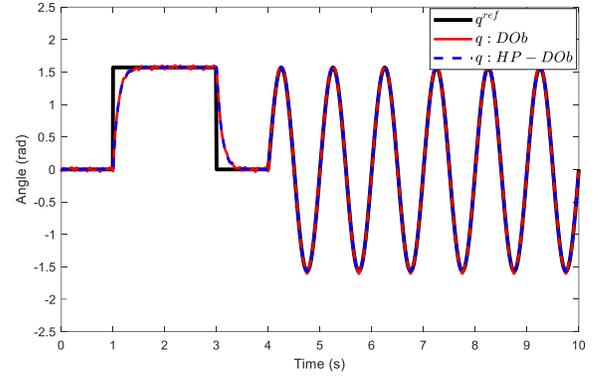

a) Position control of the servo system.

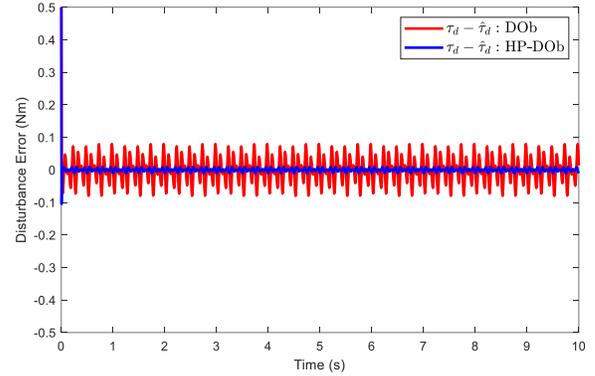

b) Disturbance estimation error

Fig. 4: Position control using PD controller and the conventional and high-performance DObs.

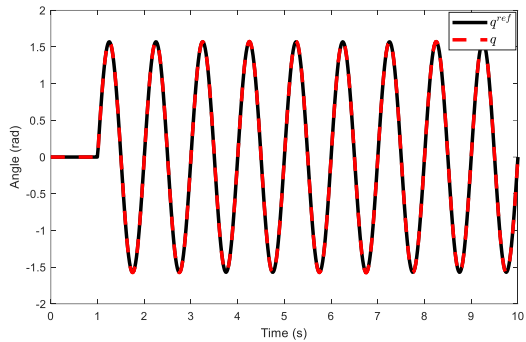

a) Position control of the servo system.

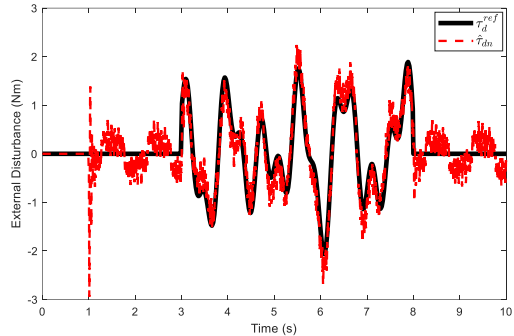

b) External disturbance and the estimated nominal disturbance

Fig. 5: Position control experiment.